\documentclass[conference]{IEEEtran}
\usepackage{amsmath,amsfonts}
\usepackage{amssymb}
\usepackage{enumerate,letltxmacro}
\usepackage{graphicx}
\usepackage{cite}
\usepackage{subfigure}
\usepackage{multirow}
\usepackage{algorithm}
\usepackage{algorithmic}
\usepackage{mathtools}
\usepackage{hyperref}
\usepackage{verbatim}

\usepackage{blindtext}

\usepackage[table,xcdraw]{xcolor}

\usepackage{array}
\newcolumntype{L}[1]{>{\raggedright\let\newline\\\arraybackslash\hspace{0pt}}m{#1}}
\newcolumntype{C}[1]{>{\centering\let\newline\\\arraybackslash\hspace{0pt}}m{#1}}
\newcolumntype{R}[1]{>{\raggedleft\let\newline\\\arraybackslash\hspace{0pt}}m{#1}}
\hypersetup{draft}

\IEEEoverridecommandlockouts

\newtheorem{theorem}{Theorem}

\newtheorem{definition}{Definition}

\newtheorem{corollary}{Corollary}
\newtheorem{proposition}{Proposition}

\newtheorem{remark}{Remark}

\newcommand{\set}[1]{\mathcal{#1}}

\newcommand{\X}{\set{X}}

\title{A Quasi-Uniform Approach to Characterizing the Boundary of the Almost Entropic Region}
\author{\IEEEauthorblockN{Satyajit Thakor and Dauood Saleem}
\IEEEauthorblockN{School of Computing and Electrical Engineering\\
Indian Institute of Technology Mandi, Himachal Pradesh, 
India}
email: satyajit@iitmandi.ac.in, D15048@students.iitmandi.ac.in
}

\begin{document}
\maketitle

\begin{abstract}
The convex closure of entropy vectors for quasi-uniform random vectors is the same as the closure of the entropy region. Thus, quasi-uniform random vectors constitute an important class of random vectors for characterizing the entropy region. Moreover, the one-to-one correspondence between quasi-uniform codes and quasi-uniform random vectors makes quasi-uniform random vectors of central importance for designing effective codes for communication systems. 
In this paper, we present a novel approach that utilizes quasi-uniform random vectors for characterizing the boundary of the almost entropic region. In particular, we use the notion of quasi-uniform random vectors to establish looseness of known inner bounds for the entropy vectors at the boundary of the almost entropic region for three random variables. For communication models such as network coding, our approach can be applied to design network codes from quasi-uniform entropy vectors.
\end{abstract}

\section{Introduction}
Quasi-uniform random vectors were introduced in \cite{Cha01}
to obtain linear information inequalities using a new combinatorial approach. It was also shown that the convex closure of the set of entropy vectors of quasi-uniform random vectors is the same as the closure of the set of all entropy vectors \cite{Han75} referred to as the almost entropic region \cite{Yeu97}. 
This result indicates that quasi-uniform random vectors constitute a very important subset of all discrete random vectors. It is known that the entropy vectors are polymatroidal \cite{Fuj78} and for three or less random variables, all polymatroids constitute the almost entropic region. However, for four or more random variables the almost entropic region is a strict subset of the set of all polymatroids \cite{ZhaYeu98}.

In \cite{ChaGraBri13}, codes induced by quasi-uniform random vectors were introduced. These codes are referred to as quasi-uniform codes and they include all linear and almost affine codes. The zero-error capacity of a class of networks was characterized in \cite{ChaGra10} by using quasi-uniform network codes. An explicit non-linear quasi-uniform code was presented in \cite{Sha11}, \cite{ShaHas10} 
for the well-known V\'{a}mos network. Thus, quasi-uniform random vectors are very important in code design. However, there is limited progress in utilizing the quasi-uniform random vectors for solving problems in coding and communications. This limited progress is mainly due to the observations that generating quasi-uniform random vectors is noted as a hard combinatorial problem \cite{ShaHas10}, and an optimization problem over quasi-uniform random vectors is known to be extremely hard to solve \cite{ChaGraBri13}. In recent work, a recursive algorithm was given to verify whether a vector is the entropy vector of some quasi-uniform random vector \cite{SalThaTiw21}. 

The capacity of various communication models is also intrinsically related to a set of ``constrained'' entropy vectors, and hence more understanding of this set can directly help in characterizing the capacity and designing optimal codes, e.g., see \cite{YanYeuZha12}, \cite{ChaGra10}. Usually, these constraints (for instance, functional dependence and conditional independence) are induced by the underlying communication model and they restrict the entropy vectors to be at the boundary of the almost entropic region. In \cite{HoChagra12}, the authors relate a problem of obtaining a converse proof for a secure communication system to characterizing a set of entropy vectors at the boundary of the almost entropic region. In \cite{TiwTha19}, the authors show that existing inner bounds for entropy vectors in specific two subsets of the boundary (of the almost entropic region for three random variables) are loose. Thus, an important step towards a complete characterization of the entropy vectors at the boundary is to establish whether the existing bounds are tight.

In this paper, we present a novel approach that utilizes the notion of quasi-uniform random vectors to establish looseness of existing inner bounds for entropy vectors at the boundary. In Section \ref{sec:Background}, we give a brief background on entropy vectors, quasi-uniform random vectors and the boundary of almost entropic region for three random variables in terms of its proper faces. In Section \ref{sec:Main Results}, we present our main results on establishing looseness of inner bounds using quasi-uniform approach. 
Finally, we discuss conclusion, future directions and implications of our main results in Section \ref{sec:Conclusion}. 

\textit{Notation}: Where no confusion will arise, singletons will be written without braces and set union will be denoted by juxtaposition. All logarithms are binary and all Shannon entropies are in bits.
\section{Background}\label{sec:Background}
\subsection{Entropy region and quasi-uniform entropy vectors}
The entropy vector of a discrete random vector $\mathbf{X}_{[n]}=(X_i,i\in [n])$, $[n]\triangleq\{1,2, \ldots,n\}$, is 
\begin{align}
\mathbf{h}\triangleq\left[h_\alpha, \emptyset\neq \alpha \subseteq [n]\right]^\intercal \in \mathbb{R}^{2^n-1}
\end{align}
where $h_\alpha$ is the Shannon entropy of 
$\mathbf{X}_{\alpha}$. The set of all entropy vectors is called the entropy region and denoted $\Gamma_n^*$\cite{Yeu97}. This region has complex structure, however, its closure $\overline{\Gamma}_n^*$ is a pointed convex cone and referred to as the almost entropic region. The set of polymatroidal vectors $\Gamma_n$ is defined as the set of vectors $\mathbf{h}$ satisfying the basic (elemental) inequalities:
\begin{align}
h_{[n]}&\geq h_{[n]\setminus i}, i\in [n] \label{eq:elemental1}\\
h_{i\beta}+h_{j\beta}&\geq h_{\beta}+h_{ij\beta}, i\neq j, \beta \subseteq [n]\setminus
\{i,j\}\label{eq:elemental2}
\end{align}			
$\Gamma_n$ is a polyhedral cone and forms an outer bound on $\overline{\Gamma}_n^*$.

The asymptotic equipartition property is a well-known notion in information theory, e.g., see \cite{Yeu08}. Quasi-uniform distributions \cite{Cha01} described below possess the non-asymptotic equipartition property.
\begin{definition}\label{def:qu_vector}
A discrete random variable $X$ with alphabet $\X$ is \textit{quasi-uniform} if $\text{Pr}\{X=x\}\in\{p,0\}$ for all $x\in \X$ where $p$ is some constant in $(0,1]$. A random vector
$\mathbf{X}_{[n]}$ is \textit{quasi-uniform} if for all $\emptyset \neq \alpha \subseteq [n]$, $\mathbf{X}_{\alpha}$ are quasi-uniform. 
\end{definition}

For a quasi-uniform random vector $\mathbf{X}_{[n]}$, it is straightforward to note that, $\text{Pr}\{\mathbf{X}_\alpha=\mathbf{x}_\alpha\}\in\{p_\alpha,0\}$ implies 
\begin{align}
p_\alpha =1/m_\alpha\label{eq:supportsize}
\end{align}
where $m_\alpha$ is the cardinality of support of $\mathbf{X}_\alpha$ and hence a natural number.
\begin{definition}
A vector $\mathbf{h}   \in  \mathbb{R}^{2^n-1}$ is a \textit{quasi-uniform entropy vector}  if there exists a quasi-uniform random vector $\mathbf{X}_{[n]}$ such that its entropy vector is $\mathbf{h}$. 
\end{definition}

The set of all quasi-uniform entropy vectors is denoted $\Lambda_n$. It was shown that the convex closure of $\Lambda_n$ is 
$\overline{\Gamma}^*_n$ \cite{Cha01}.

\subsection{Entropy vectors at the boundary of $\Gamma_3$}
It is well-known that $\Gamma^*_3 \subsetneq \overline{\Gamma}^*_3
=\Gamma_3$ and thus the relative interior of $\Gamma_3$ is a subset of $\Gamma^*_3$. Hence, the problem of characterizing $\Gamma^*_3$ reduces to characterization of entropy vectors at the boundary of $\Gamma_3$. For $n=3$, let a vector $\mathbf{h}\in \mathbb{R}^7$ be described as
\begin{align}
\mathbf{h}=[h_1,h_2,h_3,h_{12},h_{13},h_{23},h_{123}]^\intercal.
\end{align}
The polyhedral cone $\Gamma_3$ can be equivalently represented as (i) an intersection of halfspaces induced from the basic inequalities \eqref{eq:elemental1}-\eqref{eq:elemental2} and (ii) conic hull of the vectors 
\begin{align}
\mathbf{e}_1 &= [1,0,0,1,1,0,1]^\intercal, \\
\mathbf{e}_2 &= [0,1,0,1,0,1,1]^\intercal,\\
\mathbf{e}_3 &= [0,0,1,0,1,1,1]^\intercal, \\
\mathbf{e}_{12} &= [1,1,0,1,1,1,1]^\intercal,\\
\mathbf{e}_{13} &= [1,0,1,1,1,1,1]^\intercal, \\
\mathbf{e}_{23} &= [0,1,1,1,1,1,1]^\intercal,\\
\mathbf{e}_{123} &= [1,1,1,1,1,1,1]^\intercal, \\
\mathbf{e}_{123'} &= [1,1,1,2,2,2,2]^\intercal.
\end{align}
That is 
\begin{align}
    \Gamma_3 =\mathrm{cone}(\mathbf{e}_j,j\in E)
&\triangleq \left\{
\sum_{j\in E} \lambda_j\mathbf{e}_j: \lambda_j\geq 0, j\in E\right\}
\end{align}
where $E=\{1,2,3,12,13,23,123,123'\}$. The sets 
\begin{align}
    R_j \triangleq \mathrm{cone}(\mathbf{e}_j),\quad j\in E,
\end{align}
are the extreme rays of $\Gamma_3$. Only $R_{123'}$ contains non-entropy vectors and all entropy vectors in this set have the following form \cite{ZhaYeu97}.
\begin{theorem}
$\mathbf{h}\in\mathrm{cone}(\mathbf{e}_{123'})$ is an entropy vector iff
\begin{align}
    \lambda_{123'}=\log m,\quad m\in\mathbb{N}.
\end{align}
\end{theorem}

The intersection of $\Gamma_3$ with a supporting hyperplane results in a set called a proper face of $\Gamma_3$. The proper faces of $\Gamma_3$ constitute its boundary. The dimension of a face is the dimension of its affine hull. The extreme rays $R_j, j\in E$ are 1-dimensional (1-D) faces of $\Gamma_3$. For a given polyhedral cone, a face $A$ is called a subface of face $B$ if $A\subsetneq B$. We say that a vector is strictly inside a face if it is in the face and it is not in any of its subfaces, i.e., if the vector is in the relative interior of the face. For more details on polyhedral convex sets, refer \cite{Roc70}. There are 19 proper faces of $\Gamma_3$ containing non-entropy vectors (excluding faces obtained by permutation) \cite{CheYeu12}:
\begin{enumerate}[1-{D:}]
	\item  $\mathrm{cone}(\mathbf{e}_{123'})$.
	\item $\mathrm{cone}(\mathbf{e}_1, \mathbf{e}_{123'})$, $\mathrm{cone}(\mathbf{e}_{12}, \mathbf{e}_{123'})$.
	\item $\mathrm{cone}(\mathbf{e}_1,\mathbf{e}_2, \mathbf{e}_{123'})$, $\mathrm{cone}(\mathbf{e}_{12},\mathbf{e}_{13}, \mathbf{e}_{123'})$, 
	
	$\mathrm{cone}(\mathbf{e}_{1}, \mathbf{e}_{12}, \mathbf{e}_{123'})$, $\mathrm{cone}(\mathbf{e}_{1},\mathbf{e}_{23}, \mathbf{e}_{123'})$.
	\item $\mathrm{cone}(\mathbf{e}_{1}, \mathbf{e}_{2}, \mathbf{e}_{3},\mathbf{e}_{123'})$, $\mathrm{cone}(\mathbf{e}_{1},\mathbf{e}_{2}, \mathbf{e}_{12},\mathbf{e}_{123'})$, 
	
	$\mathrm{cone}(\mathbf{e}_{1}, \mathbf{e}_{2},\mathbf{e}_{13}, \mathbf{e}_{123'})$, $\mathrm{cone}(\mathbf{e}_{1},\mathbf{e}_{12}, \mathbf{e}_{13}, \mathbf{e}_{123'})$, 

	$\mathrm{cone}(\mathbf{e}_{1},\mathbf{e}_{12}, \mathbf{e}_{23},\mathbf{e}_{123'})$, $\mathrm{cone}(\mathbf{e}_{12},\mathbf{e}_{13},\mathbf{e}_{23},\mathbf{e}_{123}, \mathbf{e}_{123'})$.
	\item $\mathrm{cone}(\mathbf{e}_{1},\mathbf{e}_{2},\mathbf{e}_{3},\mathbf{e}_{12}, \mathbf{e}_{123'})$, 
	$\mathrm{cone}(\mathbf{e}_{1},\mathbf{e}_{2},\mathbf{e}_{12},\mathbf{e}_{13}, \mathbf{e}_{123'})$, $\mathrm{cone}(\mathbf{e}_{1},\mathbf{e}_{2},\mathbf{e}_{13},\mathbf{e}_{23}, \mathbf{e}_{123'})$, $\mathrm{cone}(\mathbf{e}_{1},\mathbf{e}_{12},\mathbf{e}_{13},\mathbf{e}_{23}, \mathbf{e}_{123},\mathbf{e}_{123'})$.
	\item $\mathrm{cone}(\mathbf{e}_{1},\mathbf{e}_{2},\mathbf{e}_{3},\mathbf{e}_{12}, \mathbf{e}_{13},\mathbf{e}_{123'})$, 
	
	$\mathrm{cone}(\mathbf{e}_{1},\mathbf{e}_{2},\mathbf{e}_{12},\mathbf{e}_{13},\mathbf{e}_{23}, \mathbf{e}_{123},\mathbf{e}_{123'})$.
\end{enumerate}

The entropy vectors in 2-dimensional faces are fully characterized as follows \cite{Mat06}, \cite{CheYeu12}.
\begin{theorem}
\label{thm:2-dim-Matus}
$\mathbf{h}\in$ $\mathrm{cone}(\mathbf{e}_{12}, \mathbf{e}_{123'})$ is an entropy vector iff
\begin{align}
    \lambda_{12}+\lambda_{123'}\geq \log \lceil 2^{\lambda_{123'}}\rceil.
\end{align}
\end{theorem}
\begin{theorem}
\label{thm:2-dim-Chen}
$\mathbf{h}\in$ $\mathrm{cone}(\mathbf{e}_1, \mathbf{e}_{123'})$ is an entropy vector iff
\begin{align}
\lambda_{123'}=\log m,\quad m\in\mathbb{N}.
\end{align}
\end{theorem}

A complete characterization of entropy vectors in higher dimensional faces (containing non-entropy vectors) is an open problem. However, the following property of the entropy vectors \cite{Yeu08} facilitates  formulating inner bounds for the entropy region in the proper faces.
\begin{proposition}
\label{prop:1}
		If $\mathbf{h}$ and $\mathbf{h}'$ are in  ${\Gamma}^*_n$, then $\mathbf{h} + \mathbf{h}'\in{\Gamma}^*_n$.
\end{proposition}

It was shown in \cite{TiwTha19} that the inner bounds obtainable using Proposition \ref{prop:1} and entropy vectors in lower dimensional subfaces are loose for $\mathrm{cone}(\mathbf{e}_{12},\mathbf{e}_{23}, \mathbf{e}_{123'})$ and $\mathrm{cone}(\mathbf{e}_{12},$ $\mathbf{e}_{13},\mathbf{e}_{23},\mathbf{e}_{123}, \mathbf{e}_{123'})$.
\section{Looseness of Inner Bounds for Entropy Vectors at the Boundary of $\Gamma_3$}\label{sec:Main Results}
In this section, we show using a new quasi-uniform approach that the known inner bounds for the entropy region in faces $\mathrm{cone}(\mathbf{e}_{1}, \mathbf{e}_{2}, \mathbf{e}_{3},\mathbf{e}_{123'})$ and $\mathrm{cone}(\mathbf{e}_{1}, \mathbf{e}_{2}, \mathbf{e}_{3},\mathbf{e}_{12}, \mathbf{e}_{123'})$ are loose. For a set $\Delta\subseteq \mathbb{R}^{2^n-1}$, the set $\Delta\cap\Gamma_n^*$ is referred to as the entropy region in $\Delta$.

\subsection{A 4-dimensional face}
An inner bound for the entropy region in the face 
\begin{align}
\Theta\triangleq \mathrm{cone}(\mathbf{e}_{1}, \mathbf{e}_{2}, \mathbf{e}_{3},\mathbf{e}_{123'})
\end{align}
can be obtained by Proposition \ref{prop:1} and the characterization of the entropy vectors in the subfaces $\mathrm{cone}(\mathbf{e}_{j}$, $\mathbf{e}_{123'})$, $j\in[3]$, Theorem \ref{thm:2-dim-Chen}, as follows. 

\begin{definition}
Let $\Theta^{\mathrm{in}}$ be the set of all vectors
\begin{align}\label{eq:inner1}
\mathbf{h}=\lambda_{1}\mathbf{e}_1+\lambda_{2}\mathbf{e}_2+\lambda_{3}\mathbf{e}_3+\lambda_{123'}\mathbf{e}_{123'}
\end{align}
satisfying $\lambda_j\geq 0, 
j\in[3],$ and
\begin{align}
\lambda_{123'}&=\log m, \quad m\in \mathbb{N}.\label{eq:inner1_cond}
\end{align}
\end{definition}

\begin{corollary}
The set $\Theta^{\mathrm{in}}$ is an inner bound for the entropy region in $\Theta$ (i.e., $\Theta^{\mathrm{in}}\subseteq \Theta\cap\Gamma_3^*$).
\end{corollary}
\begin{IEEEproof}
A vector $\mathbf{h}\in \Theta^{\mathrm{in}}$, of the form \eqref{eq:inner1}, is a conic combination of $\mathbf{e}_{1}, \mathbf{e}_{2}, \mathbf{e}_{3},\mathbf{e}_{123'}$ (with the coefficients $\lambda_{1},\lambda_{2},\lambda_{3},\lambda_{123'}$) and hence $\mathbf{h} \in \Theta$. By Theorem \ref{thm:2-dim-Chen}, $\mathbf{a}\triangleq \lambda_{1}\mathbf{e}_1+\lambda_{123'}\mathbf{e}_{123'}$ is an entropy vector since it satisfies \eqref{eq:inner1_cond}. 
The vector $\mathbf{a}'\triangleq\lambda_{2}\mathbf{e}_2+\lambda_{3}\mathbf{e}_3$ is also an entropy vector (since all vectors in $R_j, j\in[3]$ are entropy vectors). Hence, by Proposition \ref{prop:1}, $\mathbf{h}=\mathbf{a}+\mathbf{a}'$ is an entropy vector.
\end{IEEEproof}

\begin{theorem}\label{thm:inner1} 
There exists a quasi-uniform entropy vector strictly in $\Theta$ such that it is not in $\Theta^{\mathrm{in}}$. 
\end{theorem} 

\begin{IEEEproof}
Our approach to proving this theorem is to first construct a vector in the face such that it is a likely candidate of $\Lambda_3$. 
Then we verify whether it is (i) indeed in $\Lambda_3$ and (ii) strictly in $\Theta$ and outside the inner bound.

By \eqref{eq:supportsize}, a necessary condition for a vector $\mathbf{h}\in \overline{\Gamma}^*_n$ to be a quasi-uniform entropy vector is that $2^{h_{\alpha}}$ is a natural number for all $\emptyset\neq\alpha \subseteq [n]$. In other words, 
\begin{align}\label{eq:1}
{h_{\alpha}}=\log m_\alpha,\quad 
\forall \emptyset\neq\alpha \subseteq [n]
\end{align}
where $m_\alpha \in \mathbb{N}$.

We construct a vector in $\Theta$ satisfying \eqref{eq:1} by  choosing coefficients $\lambda_j=\log 3$, $j\in[3]$ and $\lambda_{123'}=\log (4/3)$ in \eqref{eq:inner1} as follows:
\begin{align}\label{eq:2}
\mathbf{f}
=&\log 3 \mathbf{e}_1+\log 3\mathbf{e}_2+\log 3\mathbf{e}_3+\log \frac{4}{3} \mathbf{e}_{123'}\nonumber\\
=&[\log 4,\log 4,\log 4,\log 16,\log 16,\log 16,\log 48]^\intercal
\end{align}

The vector $\mathbf{f}$ is indeed a quasi-uniform entropy vector. This is shown by an explicit quasi-uniform distribution in Table \ref{tab:1} such that $\mathbf{f}$  is its entropy vector. A joint distribution of random variables $X_1,X_2,X_3$ with the common alphabet $\mathcal X=\{a,b,c,d\}$ is given where each bullet represents the value $1/48$ and each empty box represents the value 0. It can be easily verified that $(X_i,X_j), \{i,j\}\subset [3]$, are uniformly distributed over $\mathcal X^2$ and $X_i, i\in[3]$, are uniformly distributed over $\mathcal X$. Thus, $(X_1,X_2,X_3)$ is a quasi-uniform random vector.

\begin{table}[]
	\centering
	\caption{A joint distribution 
	with its entropy vector as $\mathbf{f}$.}
	\label{tab:1}
\begin{tabular}{clcccc}
                                 & \multicolumn{1}{r}{$X_3=$}   & $a$                            & $b$                            & $c$                            & $d$                            \\ \cline{3-6} 
\multicolumn{1}{r}{$(X_1,X_2)=$} & \multicolumn{1}{l|}{$(a,a)$} & \multicolumn{1}{c|}{$\bullet$} & \multicolumn{1}{c|}{$\bullet$} & \multicolumn{1}{c|}{$\bullet$} & \multicolumn{1}{c|}{}          \\ \cline{3-6} 
                                 & \multicolumn{1}{l|}{$(a,b)$} & \multicolumn{1}{c|}{}          & \multicolumn{1}{c|}{$\bullet$} & \multicolumn{1}{c|}{$\bullet$} & \multicolumn{1}{c|}{$\bullet$} \\ \cline{3-6} 
                                 & \multicolumn{1}{l|}{$(a,c)$} & \multicolumn{1}{c|}{$\bullet$} & \multicolumn{1}{c|}{}          & \multicolumn{1}{c|}{$\bullet$} & \multicolumn{1}{c|}{$\bullet$} \\ \cline{3-6} 
                                 & \multicolumn{1}{l|}{$(a,d)$} & \multicolumn{1}{c|}{$\bullet$} & \multicolumn{1}{c|}{$\bullet$} & \multicolumn{1}{c|}{}          & \multicolumn{1}{c|}{$\bullet$} \\ \cline{3-6} 
                                 & \multicolumn{1}{l|}{$(b,a)$} & \multicolumn{1}{c|}{}          & \multicolumn{1}{c|}{$\bullet$} & \multicolumn{1}{c|}{$\bullet$} & \multicolumn{1}{c|}{$\bullet$} \\ \cline{3-6} 
                                 & \multicolumn{1}{l|}{$(b,b)$} & \multicolumn{1}{c|}{$\bullet$} & \multicolumn{1}{c|}{}          & \multicolumn{1}{c|}{$\bullet$} & \multicolumn{1}{c|}{$\bullet$} \\ \cline{3-6} 
                                 & \multicolumn{1}{l|}{$(b,c)$} & \multicolumn{1}{c|}{$\bullet$} & \multicolumn{1}{c|}{$\bullet$} & \multicolumn{1}{c|}{}          & \multicolumn{1}{c|}{$\bullet$} \\ \cline{3-6} 
                                 & \multicolumn{1}{l|}{$(b,d)$} & \multicolumn{1}{c|}{$\bullet$} & \multicolumn{1}{c|}{$\bullet$} & \multicolumn{1}{c|}{$\bullet$} & \multicolumn{1}{c|}{}          \\ \cline{3-6} 
                                 & \multicolumn{1}{l|}{$(c,a)$} & \multicolumn{1}{c|}{$\bullet$} & \multicolumn{1}{c|}{}          & \multicolumn{1}{c|}{$\bullet$} & \multicolumn{1}{c|}{$\bullet$} \\ \cline{3-6} 
                                 & \multicolumn{1}{l|}{$(c,b)$} & \multicolumn{1}{c|}{$\bullet$} & \multicolumn{1}{c|}{$\bullet$} & \multicolumn{1}{c|}{}          & \multicolumn{1}{c|}{$\bullet$} \\ \cline{3-6} 
                                 & \multicolumn{1}{l|}{$(c,c)$} & \multicolumn{1}{c|}{$\bullet$} & \multicolumn{1}{c|}{$\bullet$} & \multicolumn{1}{c|}{$\bullet$} & \multicolumn{1}{c|}{}          \\ \cline{3-6} 
                                 & \multicolumn{1}{l|}{$(c,d)$} & \multicolumn{1}{c|}{}          & \multicolumn{1}{c|}{$\bullet$} & \multicolumn{1}{c|}{$\bullet$} & \multicolumn{1}{c|}{$\bullet$} \\ \cline{3-6} 
                                 & \multicolumn{1}{l|}{$(d,a)$} & \multicolumn{1}{c|}{$\bullet$} & \multicolumn{1}{c|}{$\bullet$} & \multicolumn{1}{c|}{}          & \multicolumn{1}{c|}{$\bullet$} \\ \cline{3-6} 
                                 & \multicolumn{1}{l|}{$(d,b)$} & \multicolumn{1}{c|}{$\bullet$} & \multicolumn{1}{c|}{$\bullet$} & \multicolumn{1}{c|}{$\bullet$} & \multicolumn{1}{c|}{}          \\ \cline{3-6} 
                                 & \multicolumn{1}{l|}{$(d,c)$} & \multicolumn{1}{c|}{}          & \multicolumn{1}{c|}{$\bullet$} & \multicolumn{1}{c|}{$\bullet$} & \multicolumn{1}{c|}{$\bullet$} \\ \cline{3-6} 
                                 & \multicolumn{1}{l|}{$(d,d)$} & \multicolumn{1}{c|}{$\bullet$} & \multicolumn{1}{c|}{}          & \multicolumn{1}{c|}{$\bullet$} & \multicolumn{1}{c|}{$\bullet$} \\ \cline{3-6} 
\end{tabular}
\end{table}

It is straightforward to verify that $\mathbf{f}$  is strictly in $\Theta$ since it 
cannot be represented as a conic combination of vectors in a strict subset of $\{\mathbf{e}_{1}, \mathbf{e}_{2}, \mathbf{e}_{3},\mathbf{e}_{123'}\}$. Alternatively, it can be checked that $\mathbf{f}$ does not satisfy necessary conditions to be in any of subface of $\Theta$. For instance, each vector in the subface $\mathrm{cone}(\mathbf{e}_{1}, \mathbf{e}_{2},\mathbf{e}_{123'})$ of $\Theta$ 
must satisfy \eqref{eq:subface_cond1} but $\mathbf{f}$ violates this condition: 

For each vector $\mathbf{h}\in \mathrm{cone}(\mathbf{e}_{1}, \mathbf{e}_{2},\mathbf{e}_{123'})$, 
\begin{align}
\mathbf{h}=&\lambda_{1}\mathbf{e}_1+\lambda_{2}\mathbf{e}_2+\lambda_{123'}\mathbf{e}_{123'}\nonumber\\
=& [\lambda_1+\lambda_{123'},\lambda_2+\lambda_{123'},\lambda_{123'},\lambda_1+\lambda_2+2\lambda_{123'},\nonumber\\
&\lambda_1+2\lambda_{123'},\lambda_2+2\lambda_{123'},\lambda_1+\lambda_2+2\lambda_{123'}]^\intercal
\end{align}
and hence 
\begin{align}\label{eq:subface_cond1}
h_{12}=h_{123}.
\end{align}
However, $\mathbf{f}$ violates this condition and so $\mathbf{f}\not\in\mathrm{cone}(\mathbf{e}_{1}, \mathbf{e}_{2},\mathbf{e}_{123'})$. Similarly, it can be verified that $\mathbf{f}$ does not satisfy necessary conditions to be in other subfaces of $\Theta$.

It remains to show that $\mathbf{f}$  is not an element of $\Theta^{\mathrm{in}}$. Note that for the vector $\mathbf{f}$,
$\lambda_{123'}=\log (4/3)$
where $4/3$ is not a natural number and hence $\mathbf{f}$  violates the condition \eqref{eq:inner1_cond}.
\end{IEEEproof}

\subsection{A 5-dimensional face}
An inner bound for the entropy region in the face
\begin{align}
\Omega\triangleq \mathrm{cone}(\mathbf{e}_{1}, \mathbf{e}_{2}, \mathbf{e}_{3},\mathbf{e}_{12}, \mathbf{e}_{123'})
\end{align}
can be obtained by Proposition \ref{prop:1} and the characterization of the entropy vectors in the subfaces $\mathrm{cone}(\mathbf{e}_{i}$, $\mathbf{e}_{123'})$, $i=1,2,3$, and $\mathrm{cone}(\mathbf{e}_{12}, \mathbf{e}_{123'})$, Theorems \ref{thm:2-dim-Chen}, \ref{thm:2-dim-Matus}, as follows. 

\begin{definition}
Let $\Omega^{\mathrm{in}}$ be the set of all vectors
\begin{align}\label{eq:inner2}
\mathbf{h}=\lambda_{1}\mathbf{e}_1+\lambda_{2}\mathbf{e}_2+\lambda_{3}\mathbf{e}_3+\lambda_{12}\mathbf{e}_{12}+\lambda_{123'}\mathbf{e}_{123'},
\end{align}
$\lambda_j\geq 0, j\in \{1,2,3,12,123'\}$, such that at least one of the following conditions is satisfied:
\begin{align}
\lambda_{12}+\lambda_{123'} &\geq  \log \lceil 2^{\lambda_{123'}} \rceil,\label{eq:inner2_cond1}\\	    
\lambda_{123'}&=\log m, \quad m\in \mathbb{N}.\label{eq:inner2_cond2}
\end{align}
\end{definition}

\begin{corollary}
The set $\Omega^{\mathrm{in}}$ is an inner bound for the entropy region in $\Omega$.
\end{corollary}
\begin{IEEEproof}
A vector $\mathbf{h}\in \Omega^{\mathrm{in}}$, of the form \eqref{eq:inner2}, is a conic combination of $\mathbf{e}_{1}, \mathbf{e}_{2}, \mathbf{e}_{3},\mathbf{e}_{12},\mathbf{e}_{123'}$ (with the coefficients $\lambda_{1}$, $\lambda_{2}$, $\lambda_{3}$, $\lambda_{12}$, $\lambda_{123'}$) and hence $\mathbf{h} \in \Omega$. It remains to show that $\mathbf{h}$ is an entropy vector if it satisfies the condition \eqref{eq:inner2_cond1} (Case 1) or \eqref{eq:inner2_cond2} (Case 2).

\textit{\underline{Case 1}}: Assume that $\mathbf{h}$ satisfies \eqref{eq:inner2_cond1}. 
Then, by Theorem \ref{thm:2-dim-Matus}, $\mathbf{b}\triangleq \lambda_{12}\mathbf{e}_{12}+\lambda_{123'}\mathbf{e}_{123'}$ is an entropy vector. 
The vector $\mathbf{b}'\triangleq\lambda_{1}\mathbf{e}_1+\lambda_{2}\mathbf{e}_2+\lambda_{3}\mathbf{e}_3$ is also an entropy vector. Hence, by Proposition \ref{prop:1}, $\mathbf{h}=\mathbf{b}+\mathbf{b}'$ is an entropy vector.

\textit{\underline{Case 2}}: Assume that $\mathbf{h}$ satisfies \eqref{eq:inner2_cond2}. Then, by Theorem \ref{thm:2-dim-Chen}, $\mathbf{c}\triangleq \lambda_{1}\mathbf{e}_{1}+\lambda_{123'}\mathbf{e}_{123'}$ is an entropy vector. 
The vector $\mathbf{c}'\triangleq\lambda_{2}\mathbf{e}_2+\lambda_{3}\mathbf{e}_3+\lambda_{12}\mathbf{e}_{12}$ is also an entropy vector. Hence, by Proposition \ref{prop:1}, $\mathbf{h}=\mathbf{c}+\mathbf{c}'$ is an entropy vector.
\end{IEEEproof}

\begin{theorem}\label{thm:inner2} 
There exists an entropy vector strictly in $\Omega$ such that it is not in $\Omega^{\mathrm{in}}$.
\end{theorem} 

\begin{IEEEproof}
Similar to the approach in the proof of Theorem \ref{thm:inner1}, we construct a vector that is a likely candidate of $\Lambda_3 \cap \Omega$ satisfying \eqref{eq:1} by letting $\lambda_j=\log 4$, $j\in[3]$, $\lambda_{12}=\lambda_{123'}=\log (3/2)$ in \eqref{eq:inner1} as follows:
\begin{align}\label{eq:3}
&\log 4 \mathbf{e}_1+\log 4\mathbf{e}_2+\log 4\mathbf{e}_3+\log \frac{3}{2} \mathbf{e}_{12}+\log \frac{3}{2} \mathbf{e}_{123'}\nonumber\\
&=[\log 9,\log 9,\log 6,\log 54,\log 54,\log 54,\log 216]^\intercal
\end{align}

We could not verify whether \eqref{eq:3} is a quasi-uniform entropy vector using Algorithm 1 in \cite{SalThaTiw21} due to the large alphabet size of 486, support size of 216, and limited computational resources. However, we make attempts to obtain a consistent distribution manually, and as a result, we obtained a (non-quasi-uniform) distribution for this alphabet size such that its entropy vector differs from the vector \eqref{eq:3} only by the value of $h_{12}$, see \eqref{eq:4}. 

In Table \ref{tab:2}, a joint distribution of random variables $X_1,X_2,X_3$ with alphabets $\mathcal X_1=\mathcal X_2=\{a,b,c,d,e,f,g,h,i\}$ and $\mathcal X_3=\{a,b,c,d,e,f\}$ is given where each bullet represents the value $1/216$ and each empty box represents the value 0. It can be easily verified that $(X_1,X_3)$ and $(X_2,X_3)$ are uniformly distributed over $\mathcal X_1 \times \mathcal X_3$. However, $(X_1,X_2)$ has non-uniform distribution over its support;  we depict this distribution in Table \ref{tab:3}, where $\theta=1/216$. Thus, $\mathbf{X}_{[3]}$ with the distribution in Table \ref{tab:2} is not quasi-uniform.

\begin{table}[]
	\centering
	\caption{A joint distribution 
	with its entropy vector as $\mathbf{g}$.}\label{tab:2}
\begin{tabular}{rllllllllll}
                                 & \multicolumn{1}{r}{$X_2=$}   & $a$                            & $b$                            & $c$                            & $d$                            & $e$                            & $f$                            & $g$                            & $h$                            & $i$                            \\ \cline{3-11} 
\multicolumn{1}{r}{$(X_1,X_3)=$} & \multicolumn{1}{l|}{$(a,a)$} & \multicolumn{1}{l|}{$\bullet$} & \multicolumn{1}{l|}{$\bullet$} & \multicolumn{1}{l|}{$\bullet$} & \multicolumn{1}{l|}{$\bullet$} & \multicolumn{1}{l|}{}          & \multicolumn{1}{l|}{}          & \multicolumn{1}{l|}{}          & \multicolumn{1}{l|}{}          & \multicolumn{1}{l|}{}          \\ \cline{3-11} 
                                 & \multicolumn{1}{l|}{$(a,b)$} & \multicolumn{1}{l|}{}          & \multicolumn{1}{l|}{$\bullet$} & \multicolumn{1}{l|}{$\bullet$} & \multicolumn{1}{l|}{$\bullet$} & \multicolumn{1}{l|}{$\bullet$} & \multicolumn{1}{l|}{}          & \multicolumn{1}{l|}{}          & \multicolumn{1}{l|}{}          & \multicolumn{1}{l|}{}          \\ \cline{3-11} 
                                 & \multicolumn{1}{l|}{$(a,c)$} & \multicolumn{1}{l|}{}          & \multicolumn{1}{l|}{}          & \multicolumn{1}{l|}{$\bullet$} & \multicolumn{1}{l|}{$\bullet$} & \multicolumn{1}{l|}{$\bullet$} & \multicolumn{1}{l|}{$\bullet$} & \multicolumn{1}{l|}{}          & \multicolumn{1}{l|}{}          & \multicolumn{1}{l|}{}          \\ \cline{3-11} 
                                 & \multicolumn{1}{l|}{$(a,d)$} & \multicolumn{1}{l|}{}          & \multicolumn{1}{l|}{}          & \multicolumn{1}{l|}{}          & \multicolumn{1}{l|}{$\bullet$} & \multicolumn{1}{l|}{$\bullet$} & \multicolumn{1}{l|}{$\bullet$} & \multicolumn{1}{l|}{$\bullet$} & \multicolumn{1}{l|}{}          & \multicolumn{1}{l|}{}          \\ \cline{3-11} 
                                 & \multicolumn{1}{l|}{$(a,e)$} & \multicolumn{1}{l|}{}          & \multicolumn{1}{l|}{}          & \multicolumn{1}{l|}{}          & \multicolumn{1}{l|}{}          & \multicolumn{1}{l|}{$\bullet$} & \multicolumn{1}{l|}{$\bullet$} & \multicolumn{1}{l|}{$\bullet$} & \multicolumn{1}{l|}{$\bullet$} & \multicolumn{1}{l|}{}          \\ \cline{3-11} 
                                 & \multicolumn{1}{l|}{$(a,f)$} & \multicolumn{1}{l|}{}          & \multicolumn{1}{l|}{}          & \multicolumn{1}{l|}{}          & \multicolumn{1}{l|}{}          & \multicolumn{1}{l|}{}          & \multicolumn{1}{l|}{$\bullet$} & \multicolumn{1}{l|}{$\bullet$} & \multicolumn{1}{l|}{$\bullet$} & \multicolumn{1}{l|}{$\bullet$} \\ \cline{3-11} 
                                 & \multicolumn{1}{l|}{$(b,a)$} & \multicolumn{1}{l|}{}          & \multicolumn{1}{l|}{$\bullet$} & \multicolumn{1}{l|}{$\bullet$} & \multicolumn{1}{l|}{$\bullet$} & \multicolumn{1}{l|}{$\bullet$} & \multicolumn{1}{l|}{}          & \multicolumn{1}{l|}{}          & \multicolumn{1}{l|}{}          & \multicolumn{1}{l|}{}          \\ \cline{3-11} 
                                 & \multicolumn{1}{l|}{$(b,b)$} & \multicolumn{1}{l|}{}          & \multicolumn{1}{l|}{}          & \multicolumn{1}{l|}{$\bullet$} & \multicolumn{1}{l|}{$\bullet$} & \multicolumn{1}{l|}{$\bullet$} & \multicolumn{1}{l|}{$\bullet$} & \multicolumn{1}{l|}{}          & \multicolumn{1}{l|}{}          & \multicolumn{1}{l|}{}          \\ \cline{3-11} 
                                 & \multicolumn{1}{l|}{$(b,c)$} & \multicolumn{1}{l|}{}          & \multicolumn{1}{l|}{}          & \multicolumn{1}{l|}{}          & \multicolumn{1}{l|}{$\bullet$} & \multicolumn{1}{l|}{$\bullet$} & \multicolumn{1}{l|}{$\bullet$} & \multicolumn{1}{l|}{$\bullet$} & \multicolumn{1}{l|}{}          & \multicolumn{1}{l|}{}          \\ \cline{3-11} 
                                 & \multicolumn{1}{l|}{$(b,d)$} & \multicolumn{1}{l|}{}          & \multicolumn{1}{l|}{}          & \multicolumn{1}{l|}{}          & \multicolumn{1}{l|}{}          & \multicolumn{1}{l|}{$\bullet$} & \multicolumn{1}{l|}{$\bullet$} & \multicolumn{1}{l|}{$\bullet$} & \multicolumn{1}{l|}{$\bullet$} & \multicolumn{1}{l|}{}          \\ \cline{3-11} 
                                 & \multicolumn{1}{l|}{$(b,e)$} & \multicolumn{1}{l|}{}          & \multicolumn{1}{l|}{}          & \multicolumn{1}{l|}{}          & \multicolumn{1}{l|}{}          & \multicolumn{1}{l|}{}          & \multicolumn{1}{l|}{$\bullet$} & \multicolumn{1}{l|}{$\bullet$} & \multicolumn{1}{l|}{$\bullet$} & \multicolumn{1}{l|}{$\bullet$} \\ \cline{3-11} 
                                 & \multicolumn{1}{l|}{$(b,f)$} & \multicolumn{1}{l|}{$\bullet$} & \multicolumn{1}{l|}{}          & \multicolumn{1}{l|}{}          & \multicolumn{1}{l|}{}          & \multicolumn{1}{l|}{}          & \multicolumn{1}{l|}{}          & \multicolumn{1}{l|}{$\bullet$} & \multicolumn{1}{l|}{$\bullet$} & \multicolumn{1}{l|}{$\bullet$} \\ \cline{3-11} 
                                 & \multicolumn{1}{l|}{$(c,a)$} & \multicolumn{1}{l|}{}          & \multicolumn{1}{l|}{}          & \multicolumn{1}{l|}{$\bullet$} & \multicolumn{1}{l|}{$\bullet$} & \multicolumn{1}{l|}{$\bullet$} & \multicolumn{1}{l|}{$\bullet$} & \multicolumn{1}{l|}{}          & \multicolumn{1}{l|}{}          & \multicolumn{1}{l|}{}          \\ \cline{3-11} 
                                 & \multicolumn{1}{l|}{$(c,b)$} & \multicolumn{1}{l|}{}          & \multicolumn{1}{l|}{}          & \multicolumn{1}{l|}{}          & \multicolumn{1}{l|}{$\bullet$} & \multicolumn{1}{l|}{$\bullet$} & \multicolumn{1}{l|}{$\bullet$} & \multicolumn{1}{l|}{$\bullet$} & \multicolumn{1}{l|}{}          & \multicolumn{1}{l|}{}          \\ \cline{3-11} 
                                 & \multicolumn{1}{l|}{$(c,c)$} & \multicolumn{1}{l|}{}          & \multicolumn{1}{l|}{}          & \multicolumn{1}{l|}{}          & \multicolumn{1}{l|}{}          & \multicolumn{1}{l|}{$\bullet$} & \multicolumn{1}{l|}{$\bullet$} & \multicolumn{1}{l|}{$\bullet$} & \multicolumn{1}{l|}{$\bullet$} & \multicolumn{1}{l|}{}          \\ \cline{3-11} 
                                 & \multicolumn{1}{l|}{$(c,d)$} & \multicolumn{1}{l|}{}          & \multicolumn{1}{l|}{}          & \multicolumn{1}{l|}{}          & \multicolumn{1}{l|}{}          & \multicolumn{1}{l|}{}          & \multicolumn{1}{l|}{$\bullet$} & \multicolumn{1}{l|}{$\bullet$} & \multicolumn{1}{l|}{$\bullet$} & \multicolumn{1}{l|}{$\bullet$} \\ \cline{3-11} 
                                 & \multicolumn{1}{l|}{$(c,e)$} & \multicolumn{1}{l|}{$\bullet$} & \multicolumn{1}{l|}{}          & \multicolumn{1}{l|}{}          & \multicolumn{1}{l|}{}          & \multicolumn{1}{l|}{}          & \multicolumn{1}{l|}{}          & \multicolumn{1}{l|}{$\bullet$} & \multicolumn{1}{l|}{$\bullet$} & \multicolumn{1}{l|}{$\bullet$} \\ \cline{3-11} 
                                 & \multicolumn{1}{l|}{$(c,f)$} & \multicolumn{1}{l|}{$\bullet$} & \multicolumn{1}{l|}{$\bullet$} & \multicolumn{1}{l|}{}          & \multicolumn{1}{l|}{}          & \multicolumn{1}{l|}{}          & \multicolumn{1}{l|}{}          & \multicolumn{1}{l|}{}          & \multicolumn{1}{l|}{$\bullet$} & \multicolumn{1}{l|}{$\bullet$} \\ \cline{3-11} 
                                 & \multicolumn{1}{l|}{$(d,a)$} & \multicolumn{1}{l|}{}          & \multicolumn{1}{l|}{}          & \multicolumn{1}{l|}{}          & \multicolumn{1}{l|}{$\bullet$} & \multicolumn{1}{l|}{$\bullet$} & \multicolumn{1}{l|}{$\bullet$} & \multicolumn{1}{l|}{$\bullet$} & \multicolumn{1}{l|}{}          & \multicolumn{1}{l|}{}          \\ \cline{3-11} 
                                 & \multicolumn{1}{l|}{$(d,b)$} & \multicolumn{1}{l|}{}          & \multicolumn{1}{l|}{}          & \multicolumn{1}{l|}{}          & \multicolumn{1}{l|}{}          & \multicolumn{1}{l|}{$\bullet$} & \multicolumn{1}{l|}{$\bullet$} & \multicolumn{1}{l|}{$\bullet$} & \multicolumn{1}{l|}{$\bullet$} & \multicolumn{1}{l|}{}          \\ \cline{3-11} 
                                 & \multicolumn{1}{l|}{$(d,c)$} & \multicolumn{1}{l|}{}          & \multicolumn{1}{l|}{}          & \multicolumn{1}{l|}{}          & \multicolumn{1}{l|}{}          & \multicolumn{1}{l|}{}          & \multicolumn{1}{l|}{$\bullet$} & \multicolumn{1}{l|}{$\bullet$} & \multicolumn{1}{l|}{$\bullet$} & \multicolumn{1}{l|}{$\bullet$} \\ \cline{3-11} 
                                 & \multicolumn{1}{l|}{$(d,d)$} & \multicolumn{1}{l|}{$\bullet$} & \multicolumn{1}{l|}{}          & \multicolumn{1}{l|}{}          & \multicolumn{1}{l|}{}          & \multicolumn{1}{l|}{}          & \multicolumn{1}{l|}{}          & \multicolumn{1}{l|}{$\bullet$} & \multicolumn{1}{l|}{$\bullet$} & \multicolumn{1}{l|}{$\bullet$} \\ \cline{3-11} 
                                 & \multicolumn{1}{l|}{$(d,e)$} & \multicolumn{1}{l|}{$\bullet$} & \multicolumn{1}{l|}{$\bullet$} & \multicolumn{1}{l|}{}          & \multicolumn{1}{l|}{}          & \multicolumn{1}{l|}{}          & \multicolumn{1}{l|}{}          & \multicolumn{1}{l|}{}          & \multicolumn{1}{l|}{$\bullet$} & \multicolumn{1}{l|}{$\bullet$} \\ \cline{3-11} 
                                 & \multicolumn{1}{l|}{$(d,f)$} & \multicolumn{1}{l|}{$\bullet$} & \multicolumn{1}{l|}{$\bullet$} & \multicolumn{1}{l|}{$\bullet$} & \multicolumn{1}{l|}{}          & \multicolumn{1}{l|}{}          & \multicolumn{1}{l|}{}          & \multicolumn{1}{l|}{}          & \multicolumn{1}{l|}{}          & \multicolumn{1}{l|}{$\bullet$} \\ \cline{3-11} 
                                 & \multicolumn{1}{l|}{$(e,a)$} & \multicolumn{1}{l|}{}          & \multicolumn{1}{l|}{}          & \multicolumn{1}{l|}{}          & \multicolumn{1}{l|}{}          & \multicolumn{1}{l|}{$\bullet$} & \multicolumn{1}{l|}{$\bullet$} & \multicolumn{1}{l|}{$\bullet$} & \multicolumn{1}{l|}{$\bullet$} & \multicolumn{1}{l|}{}          \\ \cline{3-11} 
                                 & \multicolumn{1}{l|}{$(e,b)$} & \multicolumn{1}{l|}{}          & \multicolumn{1}{l|}{}          & \multicolumn{1}{l|}{}          & \multicolumn{1}{l|}{}          & \multicolumn{1}{l|}{}          & \multicolumn{1}{l|}{$\bullet$} & \multicolumn{1}{l|}{$\bullet$} & \multicolumn{1}{l|}{$\bullet$} & \multicolumn{1}{l|}{$\bullet$} \\ \cline{3-11} 
                                 & \multicolumn{1}{l|}{$(e,c)$} & \multicolumn{1}{l|}{$\bullet$} & \multicolumn{1}{l|}{}          & \multicolumn{1}{l|}{}          & \multicolumn{1}{l|}{}          & \multicolumn{1}{l|}{}          & \multicolumn{1}{l|}{}          & \multicolumn{1}{l|}{$\bullet$} & \multicolumn{1}{l|}{$\bullet$} & \multicolumn{1}{l|}{$\bullet$} \\ \cline{3-11} 
                                 & \multicolumn{1}{l|}{$(e,d)$} & \multicolumn{1}{l|}{$\bullet$} & \multicolumn{1}{l|}{$\bullet$} & \multicolumn{1}{l|}{}          & \multicolumn{1}{l|}{}          & \multicolumn{1}{l|}{}          & \multicolumn{1}{l|}{}          & \multicolumn{1}{l|}{}          & \multicolumn{1}{l|}{$\bullet$} & \multicolumn{1}{l|}{$\bullet$} \\ \cline{3-11} 
                                 & \multicolumn{1}{l|}{$(e,e)$} & \multicolumn{1}{l|}{$\bullet$} & \multicolumn{1}{l|}{$\bullet$} & \multicolumn{1}{l|}{$\bullet$} & \multicolumn{1}{l|}{}          & \multicolumn{1}{l|}{}          & \multicolumn{1}{l|}{}          & \multicolumn{1}{l|}{}          & \multicolumn{1}{l|}{}          & \multicolumn{1}{l|}{$\bullet$} \\ \cline{3-11} 
                                 & \multicolumn{1}{l|}{$(e,f)$} & \multicolumn{1}{l|}{$\bullet$} & \multicolumn{1}{l|}{$\bullet$} & \multicolumn{1}{l|}{$\bullet$} & \multicolumn{1}{l|}{$\bullet$} & \multicolumn{1}{l|}{}          & \multicolumn{1}{l|}{}          & \multicolumn{1}{l|}{}          & \multicolumn{1}{l|}{}          & \multicolumn{1}{l|}{}          \\ \cline{3-11} 
                                 & \multicolumn{1}{l|}{$(f,a)$} & \multicolumn{1}{l|}{}          & \multicolumn{1}{l|}{}          & \multicolumn{1}{l|}{}          & \multicolumn{1}{l|}{}          & \multicolumn{1}{l|}{}          & \multicolumn{1}{l|}{$\bullet$} & \multicolumn{1}{l|}{$\bullet$} & \multicolumn{1}{l|}{$\bullet$} & \multicolumn{1}{l|}{$\bullet$} \\ \cline{3-11} 
                                 & \multicolumn{1}{l|}{$(f,b)$} & \multicolumn{1}{l|}{$\bullet$} & \multicolumn{1}{l|}{}          & \multicolumn{1}{l|}{}          & \multicolumn{1}{l|}{}          & \multicolumn{1}{l|}{}          & \multicolumn{1}{l|}{}          & \multicolumn{1}{l|}{$\bullet$} & \multicolumn{1}{l|}{$\bullet$} & \multicolumn{1}{l|}{$\bullet$} \\ \cline{3-11} 
                                 & \multicolumn{1}{l|}{$(f,c)$} & \multicolumn{1}{l|}{$\bullet$} & \multicolumn{1}{l|}{$\bullet$} & \multicolumn{1}{l|}{}          & \multicolumn{1}{l|}{}          & \multicolumn{1}{l|}{}          & \multicolumn{1}{l|}{}          & \multicolumn{1}{l|}{}          & \multicolumn{1}{l|}{$\bullet$} & \multicolumn{1}{l|}{$\bullet$} \\ \cline{3-11} 
                                 & \multicolumn{1}{l|}{$(f,d)$} & \multicolumn{1}{l|}{$\bullet$} & \multicolumn{1}{l|}{$\bullet$} & \multicolumn{1}{l|}{$\bullet$} & \multicolumn{1}{l|}{}          & \multicolumn{1}{l|}{}          & \multicolumn{1}{l|}{}          & \multicolumn{1}{l|}{}          & \multicolumn{1}{l|}{}          & \multicolumn{1}{l|}{$\bullet$} \\ \cline{3-11} 
                                 & \multicolumn{1}{l|}{$(f,e)$} & \multicolumn{1}{l|}{$\bullet$} & \multicolumn{1}{l|}{$\bullet$} & \multicolumn{1}{l|}{$\bullet$} & \multicolumn{1}{l|}{$\bullet$} & \multicolumn{1}{l|}{}          & \multicolumn{1}{l|}{}          & \multicolumn{1}{l|}{}          & \multicolumn{1}{l|}{}          & \multicolumn{1}{l|}{}          \\ \cline{3-11} 
                                 & \multicolumn{1}{l|}{$(f,f)$} & \multicolumn{1}{l|}{}          & \multicolumn{1}{l|}{$\bullet$} & \multicolumn{1}{l|}{$\bullet$} & \multicolumn{1}{l|}{$\bullet$} & \multicolumn{1}{l|}{$\bullet$} & \multicolumn{1}{l|}{}          & \multicolumn{1}{l|}{}          & \multicolumn{1}{l|}{}          & \multicolumn{1}{l|}{}          \\ \cline{3-11} 
                                 & \multicolumn{1}{l|}{$(g,a)$} & \multicolumn{1}{l|}{$\bullet$} & \multicolumn{1}{l|}{}          & \multicolumn{1}{l|}{}          & \multicolumn{1}{l|}{}          & \multicolumn{1}{l|}{}          & \multicolumn{1}{l|}{}          & \multicolumn{1}{l|}{$\bullet$} & \multicolumn{1}{l|}{$\bullet$} & \multicolumn{1}{l|}{$\bullet$} \\ \cline{3-11} 
                                 & \multicolumn{1}{l|}{$(g,b)$} & \multicolumn{1}{l|}{$\bullet$} & \multicolumn{1}{l|}{$\bullet$} & \multicolumn{1}{l|}{}          & \multicolumn{1}{l|}{}          & \multicolumn{1}{l|}{}          & \multicolumn{1}{l|}{}          & \multicolumn{1}{l|}{}          & \multicolumn{1}{l|}{$\bullet$} & \multicolumn{1}{l|}{$\bullet$} \\ \cline{3-11} 
                                 & \multicolumn{1}{l|}{$(g,c)$} & \multicolumn{1}{l|}{$\bullet$} & \multicolumn{1}{l|}{$\bullet$} & \multicolumn{1}{l|}{$\bullet$} & \multicolumn{1}{l|}{}          & \multicolumn{1}{l|}{}          & \multicolumn{1}{l|}{}          & \multicolumn{1}{l|}{}          & \multicolumn{1}{l|}{}          & \multicolumn{1}{l|}{$\bullet$} \\ \cline{3-11} 
                                 & \multicolumn{1}{l|}{$(g,d)$} & \multicolumn{1}{l|}{$\bullet$} & \multicolumn{1}{l|}{$\bullet$} & \multicolumn{1}{l|}{$\bullet$} & \multicolumn{1}{l|}{$\bullet$} & \multicolumn{1}{l|}{}          & \multicolumn{1}{l|}{}          & \multicolumn{1}{l|}{}          & \multicolumn{1}{l|}{}          & \multicolumn{1}{l|}{}          \\ \cline{3-11} 
                                 & \multicolumn{1}{l|}{$(g,e)$} & \multicolumn{1}{l|}{}          & \multicolumn{1}{l|}{$\bullet$} & \multicolumn{1}{l|}{$\bullet$} & \multicolumn{1}{l|}{$\bullet$} & \multicolumn{1}{l|}{$\bullet$} & \multicolumn{1}{l|}{}          & \multicolumn{1}{l|}{}          & \multicolumn{1}{l|}{}          & \multicolumn{1}{l|}{}          \\ \cline{3-11} 
                                 & \multicolumn{1}{l|}{$(g,f)$} & \multicolumn{1}{l|}{}          & \multicolumn{1}{l|}{}          & \multicolumn{1}{l|}{$\bullet$} & \multicolumn{1}{l|}{$\bullet$} & \multicolumn{1}{l|}{$\bullet$} & \multicolumn{1}{l|}{$\bullet$} & \multicolumn{1}{l|}{}          & \multicolumn{1}{l|}{}          & \multicolumn{1}{l|}{}          \\ \cline{3-11} 
                                 & \multicolumn{1}{l|}{$(h,a)$} & \multicolumn{1}{l|}{$\bullet$} & \multicolumn{1}{l|}{$\bullet$} & \multicolumn{1}{l|}{}          & \multicolumn{1}{l|}{}          & \multicolumn{1}{l|}{}          & \multicolumn{1}{l|}{}          & \multicolumn{1}{l|}{}          & \multicolumn{1}{l|}{$\bullet$} & \multicolumn{1}{l|}{$\bullet$} \\ \cline{3-11} 
                                 & \multicolumn{1}{l|}{$(h,b)$} & \multicolumn{1}{l|}{$\bullet$} & \multicolumn{1}{l|}{$\bullet$} & \multicolumn{1}{l|}{$\bullet$} & \multicolumn{1}{l|}{}          & \multicolumn{1}{l|}{}          & \multicolumn{1}{l|}{}          & \multicolumn{1}{l|}{}          & \multicolumn{1}{l|}{}          & \multicolumn{1}{l|}{$\bullet$} \\ \cline{3-11} 
                                 & \multicolumn{1}{l|}{$(h,c)$} & \multicolumn{1}{l|}{$\bullet$} & \multicolumn{1}{l|}{$\bullet$} & \multicolumn{1}{l|}{$\bullet$} & \multicolumn{1}{l|}{$\bullet$} & \multicolumn{1}{l|}{}          & \multicolumn{1}{l|}{}          & \multicolumn{1}{l|}{}          & \multicolumn{1}{l|}{}          & \multicolumn{1}{l|}{}          \\ \cline{3-11} 
                                 & \multicolumn{1}{l|}{$(h,d)$} & \multicolumn{1}{l|}{}          & \multicolumn{1}{l|}{$\bullet$} & \multicolumn{1}{l|}{$\bullet$} & \multicolumn{1}{l|}{$\bullet$} & \multicolumn{1}{l|}{$\bullet$} & \multicolumn{1}{l|}{}          & \multicolumn{1}{l|}{}          & \multicolumn{1}{l|}{}          & \multicolumn{1}{l|}{}          \\ \cline{3-11} 
                                 & \multicolumn{1}{l|}{$(h,e)$} & \multicolumn{1}{l|}{}          & \multicolumn{1}{l|}{}          & \multicolumn{1}{l|}{$\bullet$} & \multicolumn{1}{l|}{$\bullet$} & \multicolumn{1}{l|}{$\bullet$} & \multicolumn{1}{l|}{$\bullet$} & \multicolumn{1}{l|}{}          & \multicolumn{1}{l|}{}          & \multicolumn{1}{l|}{}          \\ \cline{3-11} 
                                 & \multicolumn{1}{l|}{$(h,f)$} & \multicolumn{1}{l|}{}          & \multicolumn{1}{l|}{}          & \multicolumn{1}{l|}{}          & \multicolumn{1}{l|}{$\bullet$} & \multicolumn{1}{l|}{$\bullet$} & \multicolumn{1}{l|}{$\bullet$} & \multicolumn{1}{l|}{$\bullet$} & \multicolumn{1}{l|}{}          & \multicolumn{1}{l|}{}          \\ \cline{3-11} 
                                 & \multicolumn{1}{l|}{$(i,a)$} & \multicolumn{1}{l|}{$\bullet$} & \multicolumn{1}{l|}{$\bullet$} & \multicolumn{1}{l|}{$\bullet$} & \multicolumn{1}{l|}{}          & \multicolumn{1}{l|}{}          & \multicolumn{1}{l|}{}          & \multicolumn{1}{l|}{}          & \multicolumn{1}{l|}{}          & \multicolumn{1}{l|}{$\bullet$} \\ \cline{3-11} 
                                 & \multicolumn{1}{l|}{$(i,b)$} & \multicolumn{1}{l|}{$\bullet$} & \multicolumn{1}{l|}{$\bullet$} & \multicolumn{1}{l|}{$\bullet$} & \multicolumn{1}{l|}{$\bullet$} & \multicolumn{1}{l|}{}          & \multicolumn{1}{l|}{}          & \multicolumn{1}{l|}{}          & \multicolumn{1}{l|}{}          & \multicolumn{1}{l|}{}          \\ \cline{3-11} 
                                 & \multicolumn{1}{l|}{$(i,c)$} & \multicolumn{1}{l|}{}          & \multicolumn{1}{l|}{$\bullet$} & \multicolumn{1}{l|}{$\bullet$} & \multicolumn{1}{l|}{$\bullet$} & \multicolumn{1}{l|}{$\bullet$} & \multicolumn{1}{l|}{}          & \multicolumn{1}{l|}{}          & \multicolumn{1}{l|}{}          & \multicolumn{1}{l|}{}          \\ \cline{3-11} 
                                 & \multicolumn{1}{l|}{$(i,d)$} & \multicolumn{1}{l|}{}          & \multicolumn{1}{l|}{}          & \multicolumn{1}{l|}{$\bullet$} & \multicolumn{1}{l|}{$\bullet$} & \multicolumn{1}{l|}{$\bullet$} & \multicolumn{1}{l|}{$\bullet$} & \multicolumn{1}{l|}{}          & \multicolumn{1}{l|}{}          & \multicolumn{1}{l|}{}          \\ \cline{3-11} 
                                 & \multicolumn{1}{l|}{$(i,e)$} & \multicolumn{1}{l|}{}          & \multicolumn{1}{l|}{}          & \multicolumn{1}{l|}{}          & \multicolumn{1}{l|}{$\bullet$} & \multicolumn{1}{l|}{$\bullet$} & \multicolumn{1}{l|}{$\bullet$} & \multicolumn{1}{l|}{$\bullet$} & \multicolumn{1}{l|}{}          & \multicolumn{1}{l|}{}          \\ \cline{3-11} 
                                 & \multicolumn{1}{l|}{$(i,f)$} & \multicolumn{1}{l|}{}          & \multicolumn{1}{l|}{}          & \multicolumn{1}{l|}{}          & \multicolumn{1}{l|}{}          & \multicolumn{1}{l|}{$\bullet$} & \multicolumn{1}{l|}{$\bullet$} & \multicolumn{1}{l|}{$\bullet$} & \multicolumn{1}{l|}{$\bullet$} & \multicolumn{1}{l|}{}          \\ \cline{3-11} 
                                 &                              &                                &                                &                                &                                &                                &                                &                                &                                &                               
\end{tabular}
\end{table}

The entropy vector associated with this joint distribution is
\begin{align}\label{eq:4}
\mathbf{g}=&[\log 9,\log 9,\log 6,\log \zeta,\log 54,\log 54,\log 216]^\intercal
\end{align}
where 
$\zeta=\sqrt{54}\sqrt[4]{72}\sqrt[6]{108}\sqrt[12]{216}\approx 73.1091.$ 
It remains to show that the vector $\mathbf{g}$ is strictly in the face $\Omega$ and outside the inner bound $\Omega^{\mathrm{in}}$. 

\underline{\textit{$\mathbf{g}$ is strictly in $\Omega$}}: Note that for each vector 
$$\mathbf{h}=\lambda_{1}\mathbf{e}_1+\lambda_{2}\mathbf{e}_2+\lambda_{3}\mathbf{e}_3+\lambda_{12}\mathbf{e}_{12}+\lambda_{123'}\mathbf{e}_{123'}$$
in the face $\Omega$, we have
\begin{align}
h_1&=\lambda_1+\lambda_{12}+\lambda_{123'},\\
h_2&=\lambda_2+\lambda_{12}+\lambda_{123'},\\
h_3&=\lambda_3+\lambda_{123'},\\
h_{12}&=\lambda_1+\lambda_2+\lambda_{12}+2\lambda_{123'},\\
h_{13}&=\lambda_1+\lambda_3+\lambda_{12}+2\lambda_{123'},\\
h_{23}&=\lambda_2+\lambda_3+\lambda_{12}+2\lambda_{123'},\\
h_{123}&=\lambda_1+\lambda_2+\lambda_3+\lambda_{12}+2\lambda_{123'}.
\end{align}
For the vector $\mathbf{g}$, these conditions render
\begin{align}
\lambda_{12}&=h_1+h_2-h_{12}
= \log \left(\frac{81}{\zeta}\right),\\
\lambda_{3}&=h_{123}-h_{12} = \log \left(\frac{216}{\zeta}\right),\\
\lambda_{123'}&=h_3-\lambda_{3}
= \log \left(\frac{\zeta}{36}\right),\label{eq:123'_for_g}\\
\lambda_{1}=\lambda_2&=h_{1}-\lambda_{12}-\lambda_{123'} = \log 4.
\end{align}

Thus $\mathbf{g}$ can be represented as a conic combination of $\mathbf{e}_{1}$, $\mathbf{e}_{2}$, $\mathbf{e}_{3}$, $\mathbf{e}_{12}$, $\mathbf{e}_{123'}$:
\begin{align}
\log 4(\mathbf{e}_1+\mathbf{e}_2)+\log \frac{216}{\zeta}\mathbf{e}_3+\log \frac{81}{\zeta}\mathbf{e}_{12}+\log \frac{\zeta}{36}\mathbf{e}_{123'}\nonumber
\end{align}
and hence $\mathbf{g}\in \Omega$. 
It is straightforward to verify that $\mathbf{g}$  is  not in any of the subfaces of $\Omega$ and hence it is strictly in $\Omega$. For instance, each vector in the subface $\mathrm{cone}(\mathbf{e}_{1}, \mathbf{e}_{2},\mathbf{e}_{3},\mathbf{e}_{123'})$ of $\Omega$ must satisfy \eqref{eq:subface_cond2} but $\mathbf{g}$ violates this condition:

For each vector $\mathbf{h}\in \mathrm{cone}(\mathbf{e}_{1}, \mathbf{e}_{2},\mathbf{e}_{3},\mathbf{e}_{123'})=\Theta,$
\begin{align}
\mathbf{h}=&\lambda_{1}\mathbf{e}_1+\lambda_{2}\mathbf{e}_2+\lambda_{3}\mathbf{e}_3+\lambda_{123'}\mathbf{e}_{123'}\nonumber\\
=& [\lambda_1+\lambda_{123'},\lambda_2+\lambda_{123'},\lambda_3+\lambda_{123'},\nonumber\\
&\lambda_1+\lambda_2+2\lambda_{123'},\lambda_1+\lambda_3+2\lambda_{123'},\nonumber\\
&\lambda_2+\lambda_3+2\lambda_{123'},\lambda_1+\lambda_2+\lambda_3+2\lambda_{123'}]^\intercal
\end{align}
and hence 
\begin{align}\label{eq:subface_cond2}
h_{1}+h_{2}=h_{12}.
\end{align}

However, $\mathbf{g}$ violates this condition and so it is not in $\Theta$. 
Similarly, it can be shown that $\mathbf{g}$ does not satisfy necessary conditions to be in other subfaces of $\Omega$.

\begin{table}[]
	\centering
	\caption{The marginal distribution 
	of $(X_1,X_2)$ for the joint distribution 
	in Table \ref{tab:2}.}\label{tab:3}
	\begin{tabular}{ccrrrrrrrrr}
       & $X_2=$                   & \multicolumn{1}{l}{$a$}        & \multicolumn{1}{l}{$b$}        & \multicolumn{1}{l}{$c$}        & \multicolumn{1}{l}{$d$}        & \multicolumn{1}{l}{$e$}        & \multicolumn{1}{l}{$f$}        & \multicolumn{1}{l}{$g$}        & \multicolumn{1}{l}{$h$}        & \multicolumn{1}{l}{$i$}        \\ \cline{3-11} 
$X_1=$ & \multicolumn{1}{l|}{$a$} & \multicolumn{1}{r|}{$1\theta$} & \multicolumn{1}{r|}{$2\theta$} & \multicolumn{1}{r|}{$3\theta$} & \multicolumn{1}{r|}{$4\theta$} & \multicolumn{1}{r|}{$4\theta$} & \multicolumn{1}{r|}{$4\theta$} & \multicolumn{1}{r|}{$3\theta$} & \multicolumn{1}{r|}{$2\theta$} & \multicolumn{1}{r|}{$1\theta$} \\ \cline{3-11} 
       & \multicolumn{1}{l|}{$b$} & \multicolumn{1}{r|}{$1\theta$} & \multicolumn{1}{r|}{$1\theta$} & \multicolumn{1}{r|}{$2\theta$} & \multicolumn{1}{r|}{$3\theta$} & \multicolumn{1}{r|}{$4\theta$} & \multicolumn{1}{r|}{$4\theta$} & \multicolumn{1}{r|}{$4\theta$} & \multicolumn{1}{r|}{$3\theta$} & \multicolumn{1}{r|}{$2\theta$} \\ \cline{3-11} 
       & \multicolumn{1}{l|}{$c$} & \multicolumn{1}{r|}{$2\theta$} & \multicolumn{1}{r|}{$1\theta$} & \multicolumn{1}{r|}{$1\theta$} & \multicolumn{1}{r|}{$2\theta$} & \multicolumn{1}{r|}{$3\theta$} & \multicolumn{1}{r|}{$4\theta$} & \multicolumn{1}{r|}{$4\theta$} & \multicolumn{1}{r|}{$4\theta$} & \multicolumn{1}{r|}{$3\theta$} \\ \cline{3-11} 
       & \multicolumn{1}{l|}{$d$} & \multicolumn{1}{r|}{$3\theta$} & \multicolumn{1}{r|}{$2\theta$} & \multicolumn{1}{r|}{$1\theta$} & \multicolumn{1}{r|}{$1\theta$} & \multicolumn{1}{r|}{$2\theta$} & \multicolumn{1}{r|}{$3\theta$} & \multicolumn{1}{r|}{$4\theta$} & \multicolumn{1}{r|}{$4\theta$} & \multicolumn{1}{r|}{$4\theta$} \\ \cline{3-11} 
       & \multicolumn{1}{l|}{$e$} & \multicolumn{1}{r|}{$4\theta$} & \multicolumn{1}{r|}{$3\theta$} & \multicolumn{1}{r|}{$2\theta$} & \multicolumn{1}{r|}{$1\theta$} & \multicolumn{1}{r|}{$1\theta$} & \multicolumn{1}{r|}{$2\theta$} & \multicolumn{1}{r|}{$3\theta$} & \multicolumn{1}{r|}{$4\theta$} & \multicolumn{1}{r|}{$4\theta$} \\ \cline{3-11} 
       & \multicolumn{1}{l|}{$f$} & \multicolumn{1}{r|}{$4\theta$} & \multicolumn{1}{r|}{$4\theta$} & \multicolumn{1}{r|}{$3\theta$} & \multicolumn{1}{r|}{$2\theta$} & \multicolumn{1}{r|}{$1\theta$} & \multicolumn{1}{r|}{$1\theta$} & \multicolumn{1}{r|}{$2\theta$} & \multicolumn{1}{r|}{$3\theta$} & \multicolumn{1}{r|}{$4\theta$} \\ \cline{3-11} 
       & \multicolumn{1}{l|}{$g$} & \multicolumn{1}{r|}{$4\theta$} & \multicolumn{1}{r|}{$4\theta$} & \multicolumn{1}{r|}{$4\theta$} & \multicolumn{1}{r|}{$3\theta$} & \multicolumn{1}{r|}{$2\theta$} & \multicolumn{1}{r|}{$1\theta$} & \multicolumn{1}{r|}{$1\theta$} & \multicolumn{1}{r|}{$2\theta$} & \multicolumn{1}{r|}{$3\theta$} \\ \cline{3-11} 
       & \multicolumn{1}{l|}{$h$} & \multicolumn{1}{r|}{$3\theta$} & \multicolumn{1}{r|}{$4\theta$} & \multicolumn{1}{r|}{$4\theta$} & \multicolumn{1}{r|}{$4\theta$} & \multicolumn{1}{r|}{$3\theta$} & \multicolumn{1}{r|}{$2\theta$} & \multicolumn{1}{r|}{$1\theta$} & \multicolumn{1}{r|}{$1\theta$} & \multicolumn{1}{r|}{$2\theta$} \\ \cline{3-11} 
       & \multicolumn{1}{l|}{$i$} & \multicolumn{1}{r|}{$2\theta$} & \multicolumn{1}{r|}{$3\theta$} & \multicolumn{1}{r|}{$4\theta$} & \multicolumn{1}{r|}{$4\theta$} & \multicolumn{1}{r|}{$4\theta$} & \multicolumn{1}{r|}{$3\theta$} & \multicolumn{1}{r|}{$2\theta$} & \multicolumn{1}{r|}{$1\theta$} & \multicolumn{1}{r|}{$1\theta$} \\ \cline{3-11} 
\end{tabular}
\end{table}

\underline{\textit{$\mathbf{g} \not\in \Omega^{\mathrm{in}}$}}: 
In \eqref{eq:123'_for_g}, the number $\zeta/36$ is not a natural number and hence $\mathbf{g}$ violates the condition \eqref{eq:inner2_cond2}. Moreover,
\begin{align}
\log \lceil 2^{\lambda_{123'}} \rceil &=\log \left\lceil \frac{\zeta}{36} \right\rceil =\log 3
\end{align}
and
\begin{align}
\lambda_{12}+\lambda_{123'}&=\log \left(\frac{81}{\zeta}\frac{\zeta}{36}\right) =\log \frac{9}{4}
\end{align}
imply that $\mathbf{g}$ violates the condition \eqref{eq:inner2_cond1}. Thus, the entropy vector $\mathbf{g}$ is not an element of $\Omega^{\mathrm{in}}$.
\end{IEEEproof}
\begin{remark}
It would be interesting to establish whether \eqref{eq:3} is a quasi-uniform entropy vector. This problem motivates studying the use of structural properties of a target vector to speed-up the verification process compared to Algorithm 1 in \cite{SalThaTiw21}. For example, the vector \eqref{eq:3} (in fact, every vector in $\Omega$) has the structural properties 
$h_1+h_3=h_{13}$ and $h_2+h_3=h_{23}$.
\end{remark}
\section{Conclusion and Future Directions}
\label{sec:Conclusion}
In this paper, we presented a novel approach of utilizing quasi-uniform random vectors and their entropy vectors to show looseness of the known inner bounds for entropy vectors in the boundary of the almost entropic region. Moreover, the results demonstrate an application of the results in \cite{SalThaTiw21} to verify quasi-uniform entropy vectors and to obtain a consistent quasi-uniform distribution. One future direction is to explore utilizing the structural property of the target vector for faster verification for quasi-uniformity compared to \cite[Algorithm 1]{SalThaTiw21}. For network communication, the approach can be applied for network code construction (e.g., see \cite{AlaThaAbb21}) from given quasi-uniform entropy vectors via obtaining quasi-uniform distributions. 
\section*{Acknowledgment}
This work is supported by SERB, Department of Science and Technology, Government of India, under Core Research Grant CRG/2020/003331.

\bibliographystyle{ieeetr}

\end{document}